# Non-singular dislocation continuum theories: Strain gradient elasticity versus Peierls-Nabarro model

Markus Lazar *

Department of Physics,
Darmstadt University of Technology,
Hochschulstr. 6,
D-64289 Darmstadt, Germany

February 15, 2018

**Abstract**

Non-singular dislocation continuum theories are studied. A comparison between Peierls-Nabarro dislocations and straight dislocations in strain gradient elasticity is given. The non-singular displacement fields, non-singular stresses, plastic distortions and dislocation core shapes are analyzed and compared for the two models. The main conclusion of this study is that due to their characteristic properties, the non-singular displacement fields, non-singular stresses and dislocation core shape of screw and edge dislocations obtained in the framework of strain gradient elasticity are more realistic and physical than the corresponding fields of the Peierls-Nabarro model. Strain gradient elasticity of dislocations is a continuum dislocation theory including a weak nonlocality within the dislocation core and predicting the size and shape of the dislocation core. The dislocation core is narrower in the strain gradient elasticity dislocation model than in the Peierls-Nabarro model and more evenly distributed in two dimensions. The present analysis shows that for the modeling of the dislocation core structure the non-singular dislocation fields of strain gradient elasticity are the suitable ones.

---

*E-mail address: lazar@fkp.tu-darmstadt.de (M. Lazar).

# 1 Introduction

Dislocations play an important role in the understanding of many phenomena in solid state physics, materials science, crystallography and engineering. Dislocations are line defects producing distortions and self-stresses in an otherwise perfect crystal lattice. In particular, dislocations are the primary carrier of crystal plasticity. Whereas elasticity theory describes well the long-range elastic strain and stress of a dislocation, it breaks down in the dislocation core region and leads to a singularity at the dislocation line. Singular dislocation solutions are often circumvented by introducing an artificial core cut-off radius. This limits the applicability of the singular dislocation solutions (Volterra dislocations) to describe the region near the dislocation line and the dislocation core structure. In reality, every crystal dislocation has a dislocation core, a region where the translational crystal symmetry is broken and the dislocation core structure is different than the crystal structure of the perfect crystal. The dislocation core structure strongly influences the properties of the imperfect crystal containing dislocations. To model dislocations and especially dislocation cores in crystals, non-singular dislocation theories can be used in material science in order to avoid the singularities at the dislocation line of a Volterra dislocation. Non-singular dislocation theories are challenging alternatives to atomistic dislocation theories which can be computationally expensive and limited. Atomistic simulations use empirical interatomic potentials or *ab initio* calculations. Empirical interatomic potentials involve fitting of parameters to a predetermined database and may not be reliable in predicting the core properties. In addition, a real lattice theory of dislocations is still missing. Only the "point stress" defined on lattice sides is known [2, 10, 62]. A formal definition of the elastic distortion and dislocation density on lattice sides has never been given. Simple counting of the deranged bonds for dislocations gives the position only up to a lattice distance. The classical singular displacement fields of dislocations are usually used for inserting a dislocation and setting up the initial structure for atomistic simulations (see [3, 62]). Fitting to the elastic continuum fields far from the core is required for the position of the dislocation. Therefore, in a molecular dynamics simulation one does not really know where the dislocation is located; its location is always obtained from some interpolation scheme [20]. However, for the study of the core structure of dislocations, atomistic simulations are a useful tool. Thus, a non-singular dislocation continuum theory with "easy-to-use" displacement, distortion and strain fields can be useful for an efficient dislocation modeling.

In the 1940s, Peierls [53] and Nabarro [49] (see also [36, 37]) proposed a semi-discrete model of edge dislocations in a simple cubic lattice that incorporates the effect of lattice discreteness within a narrow region around the slip plane. The generalization towards fcc crystals was given by Leibfried and Dietze [38] and Dietze [8] (see also [59]). The Peierls-Nabarro model was the first continuum dislocation model that naturally predicts that the dislocation core should have a finite width. It considers the spreading of the dislocation core over the slip plane. Thus, the Peierls-Nabarro model provides an analytical nonlinear elastic model of dislocations including the dislocation core. The Peierls-Nabarro model takes account of the discreteness of the crystal lattice in the $y$-direction and removes the artificial singularities at the dislocation core present in the Volterra dislocation model. Consequently, the Peierls-Nabarro model has been used as a non-singular dislocation continuum model for the modeling of the dislocation core structure in crystals (e.g. [3, 57, 58, 39, 40, 13, 66]). Furthermore, Lubarda and Markenscoff [42, 43] (see also [41]) showed that the Peierls-Nabarro model may be viewed as a variable core model of the Volterra dislocation. The variable core model dislocation smears the Burgers vector, while producing on the slip plane the Peierls-Nabarro sinusoidal relation between the stress and the slip discontinuity with a variable dislocation width as a trigonometric identity unlike the original Peierls-Nabarro model where such a sinusoidal relationship was assumed a priori. Moreover, the variable core model reduces to the Peierls-Nabarro model with fixed core radius value and the far fields coincide with the Volterra dislocation ones [42]. On the other hand, a nonlocal version of the Peierls-Nabarro model where the atomic level stresses induced at the slip plane depend in a nonlocal way on the slip degrees of freedom was proposed by Miller *et al.* [44] (see also [54]). Miller *et al.* [44] showed that the locality assumption of the original Peierls-Nabarro model may be inaccurate for a dislocation whose core is narrow. This indicates that nonlocality is relevant for the modeling of narrow dislocation cores.

In the 1960s, Mindlin [45, 46] (see also [47]) proposed so-called strain gradient elasticity theories which are generalizations of elasticity theory containing the information of the crystal lattice by means of hyperstresses, additional material constants and internal characteristic lengths. Mindlin's theory of strain gradient elasticity [45, 47] is a well-suited framework to model the behavior of elastic materials at the nano-scale. Using ab initio calculations, Shodja *et al.* [61] found that the characteristic length scale parameters of Mindlin's gradient elasticity theory are in the order of Ångström [Å] for several fcc and bcc materials. Thus, as a generalization of classical elasticity, gradient elasticity becomes relevant for nano-mechanical phenomena at such length scales. The correspondence between the strain gradient theory and the atomistic structure of materials with the nearest and next nearest interatomic interactions was exhibited by Toupin and Grazis [64] (see also [1]). The discrete nature of solids is inherently incorporated in the formulations through the characteristic lengths. The capability of strain gradient theories in capturing size effects is a direct manifestation of the involvement of characteristic lengths.

Simplified versions of strain gradient elasticity [30, 31, 32, 34, 28, 29], which are particular cases of Mindlin's theories, were proposed and used for dislocation modeling. Simplified strain gradient elasticity theory with only one characteristic length scale parameter is known as strain gradient elasticity of Helmholtz type [30, 31]. Such strain gradient elasticity model is a continuum model of dislocations with core spreading. In other words, strain gradient elasticity delivers a dislocation continuum model with weak nonlocality in the dislocation core region. Non-singular fields (displacement, distortion, stress) of straight dislocations were obtained in the framework of gradient elasticity of Helmholtz type by Lazar and Maugin [30, 32], Lazar *et al.* [31] (see also [16]). Lazar and Maugin [30] were the first to use the framework of Mindlin's gradient elasticity with the remarkable outcome that both the stress and elastic strain singularities are removed and regularized at the dislocation line in contrast to the classical, singular elasticity solutions of Volterra dislocations. In addition, the strain gradient elasticity model of dislocations is a continuum dislocation model predicting a realistic dislocation core and that the dislocation core has a finite width. It considers the spreading of the dislocation core around the dislocation line. An important advantage of the non-singular dislocation solutions is that they are defined and non-singular at the dislocation core and at the dislocation line and, therefore, they can be used for the modeling of the core structure of dislocations. In non-singular dislocation continuum theories, the dislocation core radius $r_c$ is the measure of the size of the dislocation core, i.e. the region within which the displacements, distortions and stresses are unlikely to be close to the values of elasticity theory. Outside the dislocation core, the displacements, distortions, stresses of non-singular dislocation continuum theory agree with the corresponding ones of the Volterra dislocation model. Furthermore, the non-singular fields (displacement, distortion, stress) of dislocation loops were given by Lazar [27, 28, 29]. Such non-singular dislocation fields of dislocation loops obtained in strain gradient elasticity of Helmholtz type [28, 29] were successfully implemented in 3D discrete dislocation dynamics [55]. In a recent paper [56], the comparison between the non-singular stress fields of dislocations obtained in strain gradient elasticity theory and molecular statics calculations of stress fields of dislocations has been given for several materials and it shows a good agreement between the two dislocation models even in the dislocation core region.

Although there exist also some other non-singular dislocation continuum theories in the literature, the present paper is focused on strain gradient elasticity theory and the Peierls-Nabarro model. We note that the non-singular stresses of straight dislocations given by Gutkin and Aifantis [15] are "equivalent" to the non-singular stresses given by Lazar and Maugin [30]. However, the non-singular solutions of straight dislocations given by Gutkin and Aifantis [15] are based on an ad-hoc gradient modification of Hooke's law (see also [35, 30]). In addition, we do not consider the so-called non-singular continuum theory of dislocations given by Cai *et al.* [4], because no displacement fields of screw and edge dislocations have been derived in this theory. We refer to Po *et al.* [55] for a comparison between Cai's dislocation model and strain gradient elasticity theory of dislocations and to Schoeck [58] for a comparison between the non-singular stress fields of the Peierls-Nabarro model and of the model of Cai *et al.* [4]. We also do not consider in this paper Eringen's theory of nonlocal elasticity [11] since such a framework removes the classical singularities in the stress fields, whereas the elastic strain and displacement fields still possess the classical singularities. We mention also that non-singular solutions for the displacement, elastic stain and stress fields of screw and edge dislocations were given in the framework of the so-called dislocation gauge theory [9, 24, 25, 26] which is not the subject of the present paper.

This work provides comparisons between the displacement, elastic distortion, stress, plastic distortion and dislocation density fields for screw and edge dislocations in the classical Volterra model, the Peierls-Nabarro model and the strain gradient elasticity model of Helmholtz type. In detail, we compare two non-singular dislocation continuum theories, namely the Peierls-Nabarro dislocation theory and the strain gradient elasticity dislocation theory. The main aim of the paper is to compare the non-singular dislocation fields of a Peierls-Nabarro dislocation with the non-singular fields of a straight dislocation in strain gradient elasticity theory. We show which of these two dislocation continuum theories gives better and more realistic results for the modeling of dislocations in solids, especially for the dislocation core region.

The outline of this paper is the following. In Section 2, we show and compare the displacement fields, elastic distortion fields, stress fields, plastic distortion fields and the dislocation density fields of a screw dislocation in classical incompatible elasticity, Peierls-Nabarro model and incompatible strain gradient elasticity of Helmholtz type. In Section 3, we show and compare the displacement fields, elastic distortion fields, stress fields, plastic distortion fields and the dislocation density fields of an edge dislocation in classical incompatible elasticity, Peierls-Nabarro model and incompatible strain gradient elasticity of Helmholtz type. In Section 4, the most important outcomes of the comparison between the two models are given in the conclusions. The fundamentals of strain gradient elasticity of Helmholtz type are presented in Appendix A. In Appendix B, the characteristic length scales of Mindlin's strain gradient elasticity theory and strain gradient elasticity of Helmholtz type are given for several cubic materials.

## 2 Screw dislocation

In this section, we compare the dislocation fields of the classical dislocation model (Volterra model), the Peierls-Nabarro model and the strain gradient elasticity model for a screw dislocation. In the numerical analysis, we consider a screw dislocation in tungsten since tungsten is an elastically isotropic material (see Appendix B).

### 2.1 Classical solution

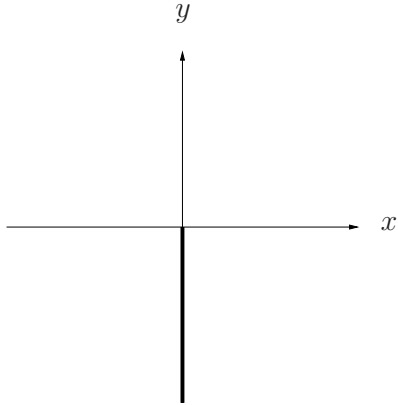

Figure 1: Branch cut for $w^0 = -\arctan(x/y)$.

Consider a screw dislocation at position $(x, y) = (0, 0)$ whose Burgers vector $b_z$ and dislocation line coincide with the direction of the $z$-axis of a Cartesian coordinate system. In the framework of classical incompatible elasticity, the discontinuous displacement field is given by

$$u_z^0 = \frac{b_z}{2\pi} w^0(x, y), \tag{1}$$

where $w^0(x, y)$ is the characteristic displacement profile function of a straight Volterra dislocation and it possesses the symmetry

$$w^0(x, y) = -w^0(-x, y). \tag{2}$$

The field $w^0$ of a Volterra screw dislocation fulfilling the condition (2) reads (see, e.g., [37])

$$w^0(x, y) = -\arctan \frac{x}{y}, \tag{3}$$

where $-\arctan \frac{x}{y}$ has the range $(-\frac{\pi}{2}, \frac{3\pi}{2})$. $w^0$ is a single-valued function with a discontinuity represented by a branch cut (see Fig. 1). The branch cut is given for $y < 0$:

$$-\arctan \frac{0^{\mp}}{y} = \pm \pi. \tag{4}$$

In the degenerate case when $y = 0$

$$-\arctan \frac{x}{y} = \begin{cases} \frac{\pi}{2}, & x < 0, \\ \text{undefined}, & x = 0, \\ -\frac{\pi}{2}, & x > 0. \end{cases} \tag{5}$$

At $y \to 0^+$: $u_z^0(x < 0, 0^+) = b_z/4$ and $u_z^0(x > 0, 0^+) = -b_z/4$. The profile of the displacement field is plotted in Fig. 2.

The total distortion, consisting of the incompatible elastic distortion and the incompatible plastic distortion, is the gradient of the displacement (1) with (3) and has two non-vanishing components in the case of a screw dislocation

$$\beta_{zx}^{\text{T},0} = \partial_x u_z^0 = -\frac{b_z}{2\pi} \left( \frac{y}{r^2} + 2\pi \, \delta(x) H(-y) \right), \tag{6}$$

$$\beta_{zy}^{\text{T},0} = \partial_y u_z^0 = \frac{b_z}{2\pi} \frac{x}{r^2}, \tag{7}$$

where $r = \sqrt{x^2 + y^2}$. Here $\delta(x)$ denotes the one-dimensional Dirac delta function and $H(y)$ is the Heaviside step function defined by

$$H(y) = \begin{cases} 0, & y < 0, \\ 1, & y > 0. \end{cases} \tag{8}$$

Because the last term in Eq. (6) is discontinuous and singular at the branch cut, it may be identified with the plastic distortion

$$\beta_{zx}^{\text{P},0} = -b_z \, \delta(x) H(-y). \tag{9}$$

This plastic distortion gives rise to a dislocation density of the screw dislocation according to

$$\alpha_{zz}^0 = \partial_y \beta_{zx}^{\text{P},0} = b_z \, \delta(x) \delta(y). \tag{10}$$

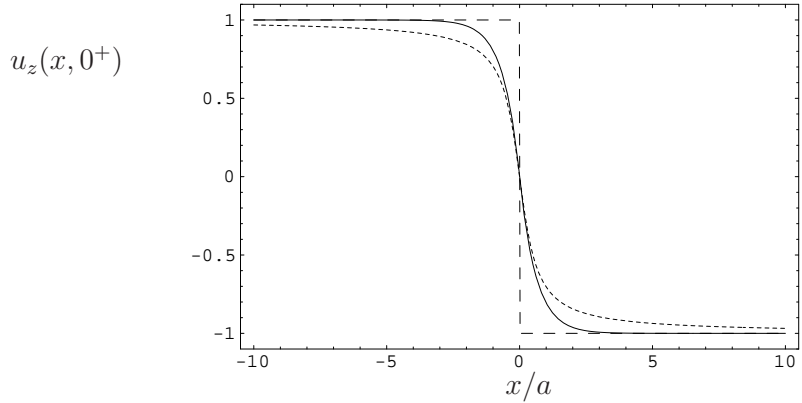

Figure 2: Displacement field near the dislocation line of a screw dislocation within strain gradient elasticity for $\ell = 0.61a$ (solid line), the Peierls-Nabarro model for $d = a$ (small dashed line), and the Volterra model (dashed line) when $y \to 0^+$ are given in units of $b_z/4$.

It means that the dislocation is concentrated at the dislocation line. Thus, the dislocation density of a Volterra dislocation is a singular Dirac $\delta$-peak at the dislocation line $(x = 0, y = 0)$.

Moreover, the stresses of the Volterra screw dislocation are [7]

$$\sigma_{zx}^0 = -\frac{\mu b_z}{2\pi} \frac{y}{r^2}, \tag{11}$$

$$\sigma_{zy}^0 = \frac{\mu b_z}{2\pi} \frac{x}{r^2}, \tag{12}$$

where $\mu$ is the shear modulus. These fields are singular at the dislocation line ($x = 0, y = 0$).

## 2.2 Peierls-Nabarro solution

Let us now consider the corresponding fields of a screw dislocation within the Peierls-Nabarro dislocation model. The Peierls-Nabarro dislocation model accounts for the discreteness of the crystalline lattice. It considers the spreading of the dislocation core in the $y = 0$ plane (slip plane). In the Peierls-Nabarro model, the crystal is separated into two half spaces, one above and one below the slip plane at $y = 0$. The dislocation core is considered as distribution of mismatch across the slip plane. The half-width of a Peierls-Nabarro dislocation is the distance where the displacement has reached half its maximum value. Note that in the variable core model [42, 43], such a separation of the crystal into two elastic half-spaces by the distance $d$ is not necessary and therefore the width or size of a dislocation can be considered as independent parameter in such a framework.

The displacement field of the Peierls-Nabarro screw dislocation at position $(0, 0)$ is

given by [37, 18]

$$u_z^{\text{PN}} = -\frac{b_z}{2\pi} \arctan \frac{x}{y \pm \eta}, \qquad (13)$$

where $\eta = d/2$ is the half-width of the Peierls-Nabarro screw dislocation core and $d$ is the interplanar spacing across the plane $y = 0$. The plus sign in $y \pm \eta$ is taken in the half-plane $y > 0$ and the minus sign is taken in the half-plane $y < 0$. In the limit $y \to 0^+$, the displacement discontinuity (13) is

$$u_z^{\text{PN}}(x, y \to 0^+) = -\frac{b_z}{2\pi} \arctan \frac{x}{\eta}. \qquad (14)$$

The profile of the displacement field (14) is plotted in Fig. 2 for $d = a$, where $a$ is the lattice parameter. Eq. (14) gives a dislocation core spreading in the $x$-direction with a wide dislocation core due to the two tails approaching the profile of a Volterra dislocation at about $10a$. Note that the displacement profile (14) corresponds to a localized "kink" or "soliton" centered about $x = 0$, which takes $u_z^{\text{PN}}(x = \infty, y \to 0^+) = -u_z^{\text{PN}}(x = -\infty, y \to 0^+) = -b_z/4$.

The non-vanishing components of the stress of a Peierls-Nabarro screw dislocation are (see, e.g., [23, 18])

$$\sigma_{zx}^{\text{PN}} = -\frac{\mu b_z}{2\pi} \frac{y \pm \eta}{x^2 + (y \pm \eta)^2}, \qquad (15)$$

$$\sigma_{zy}^{\text{PN}} = \frac{\mu b_z}{2\pi} \frac{x}{x^2 + (y \pm \eta)^2}. \qquad (16)$$

Note that there is no singularity in the displacement, strains, and stresses; however discontinuities are present in the displacement, strains, and stresses (see Figs. 2–4). The stress (15) is discontinuous across the $y = 0$ plane, namely

$$\sigma_{zx}^{\text{PN}}(x, y \to 0^\pm) = \mp \frac{\mu b_z}{2\pi} \frac{\eta}{x^2 + \eta^2} \qquad (17)$$

and at the dislocation line

$$\sigma_{zx}^{\text{PN}}(0, y \to 0^\pm) = \mp \frac{\mu b_z}{2\pi \eta}. \qquad (18)$$

The stress $\sigma_{zy}^{\text{PN}}$ has its extreme value $|\sigma_{zy}^{\text{PN}}(x, 0)| = \frac{\mu b_z}{4\pi \eta}$ at $|x| = \eta$. For $\eta = a/2$: $\sigma_{zy}^{\text{PN}}$ has its extreme value $|\sigma_{zy}^{\text{PN}}(x, 0)| = \frac{\mu b_z}{2\pi a}$ at $|x| = a/2$, and $\sigma_{zx}^{\text{PN}}(0, 0^\pm) = \mp \frac{\mu b_z}{\pi a}$. Moreover, the stresses $\sigma_{zx}^{\text{PN}}$ and $\sigma_{zy}^{\text{PN}}$ become very large in the core region and at the dislocation line (see Fig. 3). The stresses are plotted in Fig. 4 for $d = a$.

Following the so-called Eshelby smearing (ramp-core) technique [12] (see also [52, 67]), we replace the $\delta(x)$-function in the classical plastic distortion (9) by the $\delta$-sequence $f_\eta(x)$:

$$f_\eta(x) = \frac{1}{\pi} \frac{\eta}{x^2 + \eta^2} \qquad (19)$$

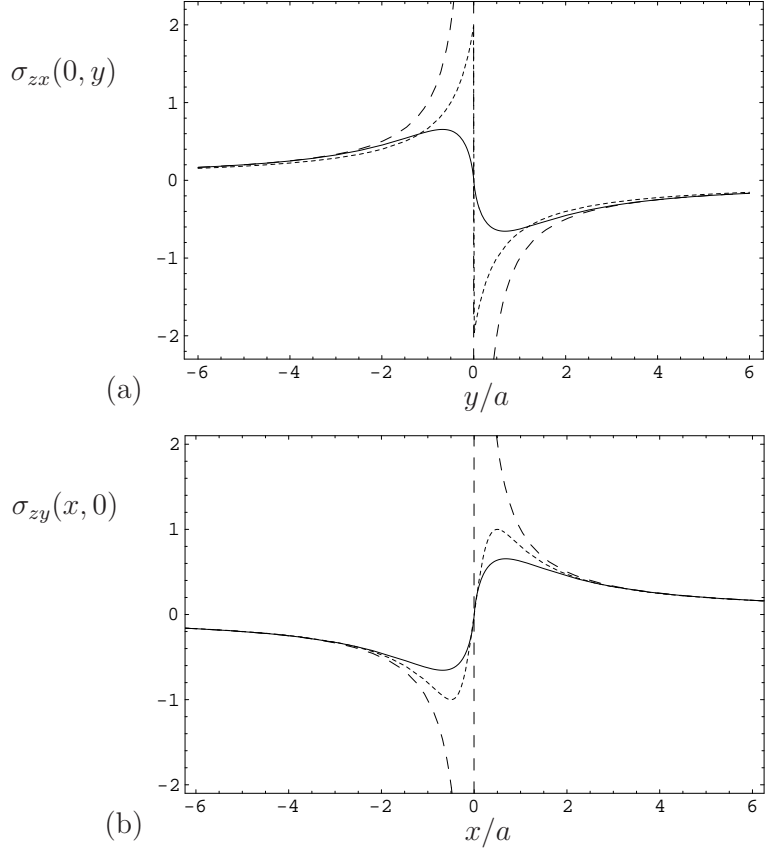

Figure 3: Stress fields of a screw dislocation: strain gradient elasticity model for $\ell = 0.61a$ (solid line), Peierls-Nabarro model for $d = a$ (small dashed line), and Volterra model (dashed line) are given in units of $\mu b_z/[2\pi a]$.

and obtain for the plastic distortion of a Peierls-Nabarro screw dislocation

$$\beta_{zx}^{\mathrm{P,PN}} = \beta_{zx}^{\mathrm{P,0}} * f_\eta = -b_z f_\eta(x) H(-y) = -\frac{b_z}{\pi} \frac{\eta}{x^2 + \eta^2} H(-y), \tag{20}$$

where $*$ denotes the one-dimensional convolution in the variable $x$. Unlike the classical plastic distortion which is singular at $x = 0$ due to $\delta(x)$ in Eq. (9), Eq. (20) is smooth there (see Fig. 6). The plastic distortion (20) possesses a minimum at the dislocation line

$$\beta_{zx}^{\mathrm{P,PN}}(0, y < 0) = -\frac{b_z}{\pi \eta}. \tag{21}$$

The corresponding dislocation density of a Peierls-Nabarro screw dislocation is

$$\alpha_{zz}^{\mathrm{PN}} = \partial_y \beta_{zx}^{\mathrm{P,PN}} = b_z f_\eta(x) \delta(y) = \frac{b_z}{\pi} \frac{\eta}{x^2 + \eta^2} \delta(y). \tag{22}$$

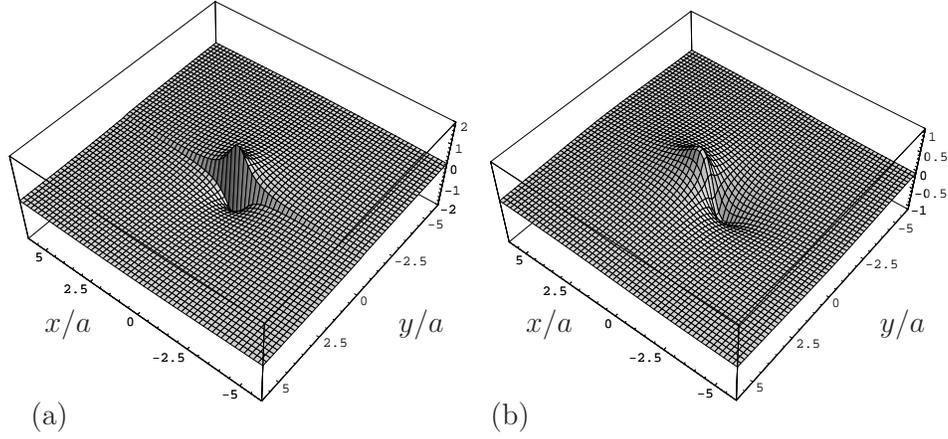

Figure 4: Stress fields of a Peierls-Nabarro screw dislocation ($d = a$) near the dislocation line: (a) $\sigma_{zx}^{\text{PN}}$ and (b) $\sigma_{zy}^{\text{PN}}$ are given in units of $\mu b_z/[2\pi a]$ .

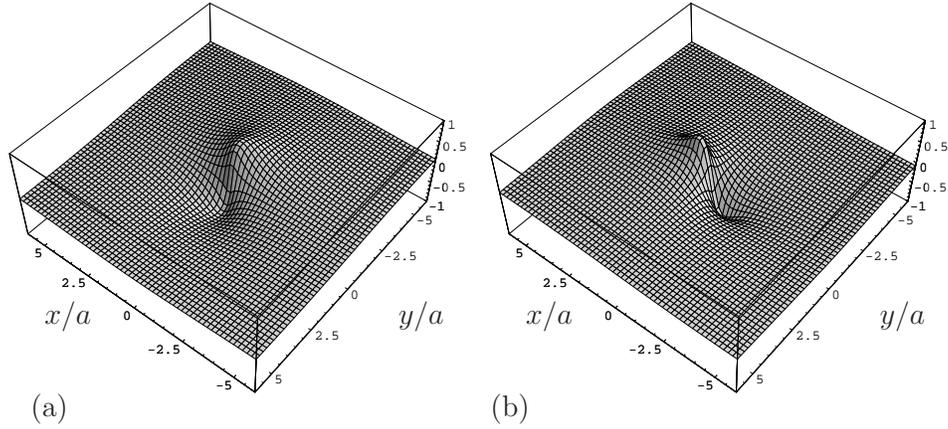

Figure 5: Stress fields of a screw dislocation in strain gradient elasticity for $\ell = 0.61 a$ near the dislocation line: (a) $\sigma_{zx}$ and (b) $\sigma_{zy}$ are given in units of $\mu b_z/[2\pi a]$ .

It can be seen that the relation $\alpha_{zz}^{\text{PN}} = \alpha_{zz}^0 * f_\eta$ holds. In comparison with the classical plastic distortion (9) and the classical dislocation density (10), the Peierls-Nabarro model regularizes only in the $x$-variable. Thus, the Peierls-Nabarro screw dislocation is smeared out in the $x$-direction and is non-zero at $y = 0$, due to the Delta-function $\delta(y)$. In Eq. (22), it can be seen that the dislocation density of a Peierls-Nabarro screw dislocation is still singular, due to the Delta-function $\delta(y)$. Such a dislocation core is usually called ramp-core or flat-core (e.g. [52]). A ramp-core is not realistic since a dislocation core is extended in reality in two-dimensions (see, e.g., [21]).

## 2.3 Gradient solution

Let us now consider the corresponding fields of a screw dislocation within the theory of strain gradient elasticity of Helmholtz type given by Lazar and Maugin [30, 32]. In strain gradient elasticity, the displacement field of a screw dislocation is

$$u_z = \frac{b_z}{2\pi} w(x,y), \qquad (23)$$

where the displacement profile function $w(x,y)$ is given by

$$w = w^0 + \int_0^\infty \frac{s \sin(sx)}{s^2 + \frac{1}{\ell^2}} \left[ \text{sgn}(y)\, e^{-|y|\sqrt{s^2 + \frac{1}{\ell^2}}} + 2H(-y) \right] ds, \qquad (24)$$

using Fourier transform [32] (see also [5]). Here $\ell$ is the characteristic length of strain gradient elasticity of Helmholtz type and $\text{sgn}(y)$ denotes the signum function defined by

$$\text{sgn}(y) = \begin{cases} -1, & y < 0, \\ 1, & y > 0, \end{cases} \qquad (25)$$

or in terms of the Heaviside function

$$\text{sgn}(y) = 2H(y) - 1. \qquad (26)$$

When $y \to 0^+$, the displacement field (24) takes the form:

$$w(x, y \to 0^+) = -\frac{\pi}{2} \text{sgn}(x) \left\{ 1 - e^{-|x|/\ell} \right\}. \qquad (27)$$

The gradient term appearing in Eq. (27) leads to a smoothing of the displacement profile unlike the jump occurring in the classical solution (see Fig. 2). The smoothing depends on the length scale $\ell$. The asymptotic limits are

$$u_z(x, 0^+) = \frac{b_z}{4} \text{ as } x \to -\infty \qquad \text{and} \qquad u_z(x, 0^+) = -\frac{b_z}{4} \text{ as } x \to \infty. \qquad (28)$$

With this smoothing of the displacement profile the dislocation core radius can be defined in terms of the length scale $\ell$ and is about $5\ell$. In the displacement profile (Fig. 2), it can be seen that the dislocation core of a screw dislocation in gradient elasticity is narrower than the core of a Peierls-Nabarro dislocation. For the chosen value of $\ell = 0.61a$ (W), the dislocation core radius is about $3a$ (see Fig. 2).

The incompatible elastic distortions read [32, 28]

$$\beta_{zx} = -\frac{b_z}{2\pi} \frac{y}{r^2} \left\{ 1 - \frac{r}{\ell} K_1(r/\ell) \right\}, \qquad (29)$$

$$\beta_{zy} = \frac{b_z}{2\pi} \frac{x}{r^2} \left\{ 1 - \frac{r}{\ell} K_1(r/\ell) \right\}, \qquad (30)$$

which are non-singular.

The stress components of a screw dislocation in strain gradient elasticity are [30, 15]

$$\sigma_{zx} = -\frac{\mu b_z}{2\pi} \frac{y}{r^2} \left\{ 1 - \frac{r}{\ell} K_1(r/\ell) \right\}, \qquad (31)$$

$$\sigma_{zy} = \frac{\mu b_z}{2\pi} \frac{x}{r^2} \left\{ 1 - \frac{r}{\ell} K_1(r/\ell) \right\}, \qquad (32)$$

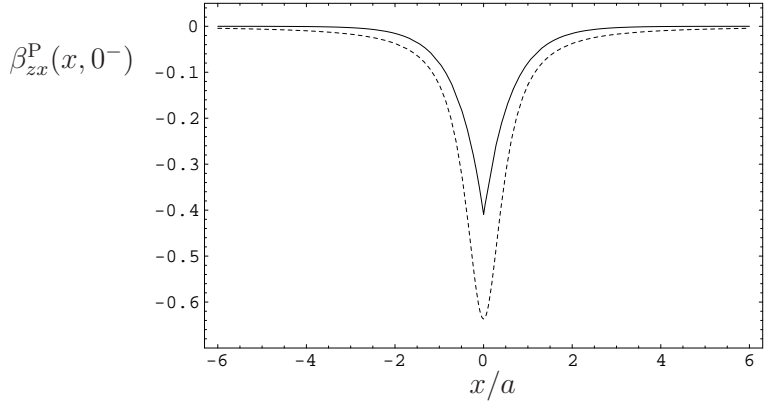

Figure 6: Plastic distortion of a screw dislocation in strain gradient elasticity for $\ell = 0.61a$ (solid line) and Peierls-Nabarro model for $\eta = a/2$ (small dashed line) given in units of $b_z/a$.

where $K_n$ denotes the modified Bessel function of order $n$. The appearance of the modified Bessel function $K_1$ in Eqs. (29) and (32) leads to the regularization of the classical singularity of order $\mathcal{O}(1/r)$ at the dislocation line and gives non-singular stresses and non-singular elastic distortions. The non-singular stresses are zero at the dislocation line. The stress $\sigma_{zy}$ has its extreme value $|\sigma_{zy}(x,0)| \simeq 0.399\frac{\mu b_z}{2\pi \ell}$ at $|x| \simeq 1.114\ell$, whereas the stress $\sigma_{zx}$ has its extreme value $|\sigma_{zx}(0,y)| \simeq 0.399\frac{\mu b_z}{2\pi \ell}$ at $|y| \simeq 1.114\ell$. For $W$ with $\ell = 0.61a$ (see Appendix B): $\sigma_{zy}$ has its extreme value $|\sigma_{zy}(x,0)| \simeq 0.654\frac{\mu b_z}{2\pi a}$ at $|x| \simeq 0.68a$, whereas $\sigma_{zx}$ has its extreme value $|\sigma_{zx}(0,y)| \simeq 0.654\frac{\mu b_z}{2\pi a}$ at $|y| \simeq 0.68a$. The non-singular stresses are plotted in Figs. 3 and 4.

In strain gradient elasticity, the plastic distortion tensor is given by

$$\beta_{zx}^{\mathrm{P}} = -\frac{b_z}{2\pi} \int_0^\infty \frac{\cos(sx)}{1+\ell^2 s^2} \left[ \mathrm{sgn}(y)\, \mathrm{e}^{-|y|\sqrt{s^2+\frac{1}{\ell^2}}} + 2H(-y) \right] \mathrm{d}s \,. \tag{33}$$

When $y \to 0$, the plastic distortion (33) takes the form:

$$\beta_{zx}^{\mathrm{P}}(x, y \to 0) = -\frac{b_z}{2\pi}\frac{\pi}{2}\frac{1}{\ell}\,\mathrm{e}^{-|x|/\ell}\,. \tag{34}$$

Unlike the classical plastic distortion which is singular at $x = 0$ due to $\delta(x)$ in Eq. (9), Eq. (34) is smooth there (see Fig. 6). The plastic distortion (34) possesses a minimum at the dislocation line

$$\beta_{zx}^{\mathrm{P}}(0,0) = -\frac{b_z}{4\ell}\,. \tag{35}$$

The corresponding dislocation density of a screw dislocation is [32, 31]

$$\alpha_{zz} = \frac{b_z}{2\pi\ell^2}\, K_0(r/\ell)\,. \tag{36}$$

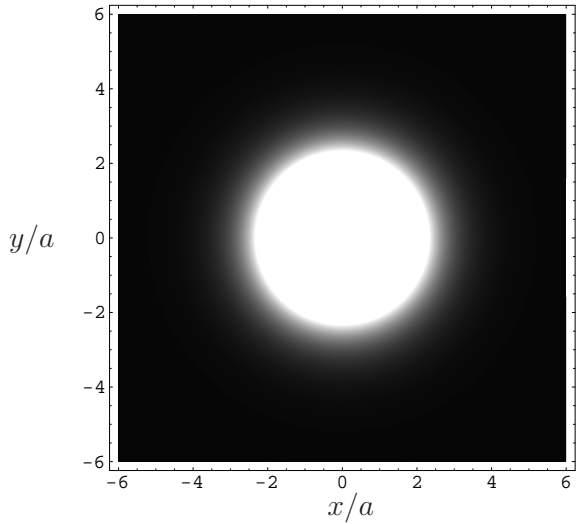

Figure 7: Contour of the dislocation density of a screw dislocation in strain gradient elasticity for $\ell = 0.61a$.

In general, the dislocation density tensor defines the dislocation core region and determines the shape and size of the dislocation core. For that reason, one could call $\alpha_{ij}$ the dislocation core tensor. Eq. (36) models the dislocation core region as a disk in the $xy$-plane in a realistic manner. The dislocation density tensor (36) is only non-zero in the dislocation core. Therefore, Eq. (36) obtained in strain gradient elasticity describes a dislocation core spreading (see Fig. 7) and represents the weak nonlocality present in the dislocation core region. Fig. 7 shows that the dislocation core radius is obtained as $2.5a \leq r_\mathrm{c} \leq 3a$ for W. However, the dislocation density (36) obtained in strain gradient elasticity of Helmholtz type contains a logarithmic singularity at the dislocation line due to the near-field behavior of the modified Bessel function, $K_0(r/\ell) \simeq -\bigl[\ln \frac{r}{2\ell} + \gamma\bigr]$, where $\gamma$ denotes the Euler constant. For a non-singular dislocation density of a straight dislocation, strain gradient elasticity of bi-Helmholtz type can be used as shown in [32, 34].

## 3 Edge dislocation

In this section, we compare the dislocation fields of the classical dislocation model (Volterra model), the Peierls-Nabarro model and the strain gradient elasticity model for an edge dislocation. In the numerical analysis, we consider an edge dislocation in tungsten.

### 3.1 Classical solution

We consider a Volterra edge dislocation at position $(0, 0)$ whose Burgers vector is parallel to the $x$-axis and the dislocation line coincides with the $z$-axis of the Cartesian coordinate system. We use the classical displacements with the branch cut at $x=0$ and for $y<0$ given by Nabarro [50, 51]

$$u_x^0 = \frac{b_x}{2\pi}\left(w^0(x,y) + \frac{xy}{2(1-\nu)r^2}\right),\tag{37}$$

$$u_y^0 = -\frac{b_x}{4\pi(1-\nu)}\left((1-2\nu)\ln r - \frac{y^2}{r^2}\right),\tag{38}$$

where the displacement profile function $w^0(x,y)$ is given in Eq. (3) and $\nu$ is the Poisson ratio. Note that only $w^0(x,y)$ is discontinuous due to the jump. All other parts of the displacements (37) and (38) are continuous. In addition, the first part of Eq. (38) has a logarithmic singularity. The non-vanishing components of the elastic distortion are [7]

$$\beta_{xx}^0 = -\frac{b_x}{4\pi(1-\nu)}\frac{y}{r^2}\left\{(1-2\nu) + \frac{2x^2}{r^2}\right\},\tag{39}$$

$$\beta_{xy}^0 = \frac{b_x}{4\pi(1-\nu)}\frac{x}{r^2}\left\{(3-2\nu) - \frac{2y^2}{r^2}\right\},\tag{40}$$

$$\beta_{yx}^0 = -\frac{b_x}{4\pi(1-\nu)}\frac{x}{r^2}\left\{(1-2\nu) + \frac{2y^2}{r^2}\right\},\tag{41}$$

$$\beta_{yy}^0 = -\frac{b_x}{4\pi(1-\nu)}\frac{y}{r^2}\left\{(1-2\nu) - \frac{2x^2}{r^2}\right\},\tag{42}$$

and the stress components are

$$\sigma_{xx}^0 = -\frac{\mu b_x}{2\pi(1-\nu)}\frac{y}{r^4}\left(y^2 + 3x^2\right),\tag{43}$$

$$\sigma_{yy}^0 = -\frac{\mu b_x}{2\pi(1-\nu)}\frac{y}{r^4}\left(y^2 - x^2\right),\tag{44}$$

$$\sigma_{xy}^0 = \frac{\mu b_x}{2\pi(1-\nu)}\frac{x}{r^4}\left(x^2 - y^2\right),\tag{45}$$

$$\sigma_{zz}^0 = -\frac{\mu b_x \nu}{\pi(1-\nu)}\frac{y}{r^2},\tag{46}$$

which are singular at the dislocation line $(x=0, y=0)$.

In addition, the plastic distortion reads

$$\beta_{xx}^{0,\mathrm{P}} = -b_x\,\delta(x)H(-y),\tag{47}$$

which is caused by the jump of $w^0(x,y)$. The dislocation density of a single edge dislocation located at the position $(0,0)$ has the following non-vanishing component (see, e.g., [7])

$$\alpha_{xz}^0 = \partial_y \beta_{xx}^{0,\mathrm{P}} = b_x\,\delta(x)\delta(y)\,.\tag{48}$$

These fields are singular at the dislocation line $(x=0, y=0)$.

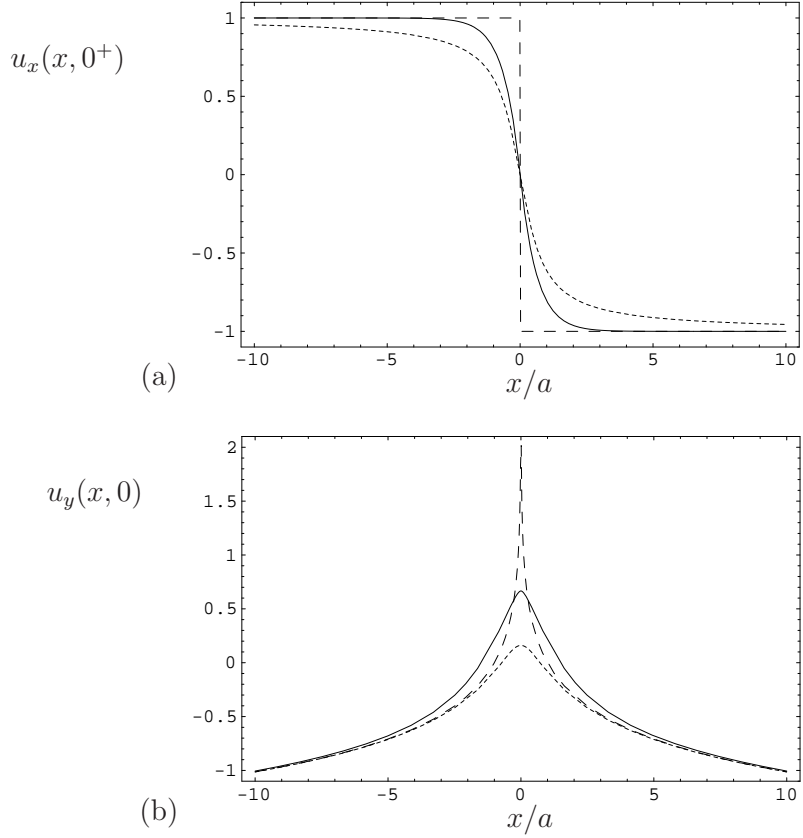

Figure 8: Displacement fields of an edge dislocation in strain gradient elasticity for $\ell = 0.61a$ (solid line), Peierls-Nabarro model for $d = a$ (small dashed line), and Volterra model (dashed line): a) $u_x$ is given in units of $b_x/4$, and b) $u_y$ is given in units of $b_x/[4\pi(1-\nu)]$ for $\nu = 0.28$.

## 3.2 Peierls-Nabarro solution

Let us consider the corresponding fields of an edge dislocation within the Peierls-Nabarro dislocation model. For a Peierls-Nabarro edge dislocation, the displacement fields are given by [36, 51, 43]

$$u_x^{\text{PN}} = -\frac{b_x}{2\pi}\left(\arctan\frac{x}{y \pm \zeta} - \frac{1}{2(1-\nu)}\frac{xy}{x^2+(y\pm\zeta)^2}\right), \tag{49}$$

$$u_y^{\text{PN}} = -\frac{b_x}{4\pi(1-\nu)}\left((1-2\nu)\ln\sqrt{\frac{x^2+(y\pm\zeta)^2}{b_x^2}} - \frac{y(y\pm\zeta)}{x^2+(y\pm\zeta)^2}\right), \tag{50}$$

where $\zeta = d/[2(1 - \nu)]$ is the half-width of the Peierls-Nabarro edge dislocation core. In the limit $y \to 0^+$, the displacements (49) and (50) become

$$u_x^{\text{PN}}(x, y \to 0^+) = -\frac{b_x}{2\pi} \arctan \frac{x}{\zeta}, \tag{51}$$

$$u_y^{\text{PN}}(x, y \to 0^+) = -\frac{b_x(1 - 2\nu)}{4\pi(1 - \nu)} \ln \sqrt{\frac{x^2 + \zeta^2}{b_x^2}}. \tag{52}$$

fulfilling the condition $u_x^{\text{PN}}(x = \infty, y \to 0^+) = -u_x^{\text{PN}}(x = -\infty, y \to 0^+) = -b_x/4$. The displacement fields of a Peierls-Nabarro edge dislocation are plotted in Fig. 8 for $d = a$. It can be seen that they are non-singular and that Eq. (51) represents the displacement discontinuity along the slip plane.

The non-vanishing stress components of the Peierls-Nabarro edge dislocation are [23, 18, 41, 43]

$$\sigma_{xx}^{\text{PN}} = -\frac{\mu b_x}{2\pi(1 - \nu)} \left\{ \frac{y \pm 2\zeta}{x^2 + (y \pm \zeta)^2} + \frac{2x^2 y}{[x^2 + (y \pm \zeta)^2]^2} \right\}, \tag{53}$$

$$\sigma_{yy}^{\text{PN}} = -\frac{\mu b_x}{2\pi(1 - \nu)} \left\{ \frac{y}{x^2 + (y \pm \zeta)^2} - \frac{2x^2 y}{[x^2 + (y \pm \zeta)^2]^2} \right\}, \tag{54}$$

$$\sigma_{xy}^{\text{PN}} = \frac{\mu b_x}{2\pi(1 - \nu)} \left\{ \frac{x}{x^2 + (y \pm \zeta)^2} - \frac{2xy(y \pm \zeta)}{[x^2 + (y \pm \zeta)^2]^2} \right\}, \tag{55}$$

$$\sigma_{zz}^{\text{PN}} = -\frac{\mu b_x \nu}{\pi(1 - \nu)} \frac{y \pm \zeta}{x^2 + (y \pm \zeta)^2}. \tag{56}$$

The stress components (53)–(56) are plotted in Figs. 9 and 10. One can see that the stresses of a Peierls-Nabarro edge dislocation are non-singular. However, the stresses (53) and (56) are discontinuous across the $y = 0$ plane, namely

$$\sigma_{xx}^{\text{PN}}(x, y \to 0^{\pm}) = \mp \frac{\mu b_x}{2\pi(1 - \nu)} \frac{2\zeta}{x^2 + \zeta^2}, \quad \sigma_{zz}^{\text{PN}}(x, y \to 0^{\pm}) = \mp \frac{\mu b_x \nu}{\pi(1 - \nu)} \frac{\zeta}{x^2 + \zeta^2} \tag{57}$$

and at the dislocation line

$$\sigma_{xx}^{\text{PN}}(0, y \to 0^{\pm}) = \mp \frac{\mu b_x}{\pi(1 - \nu)\zeta}, \quad \sigma_{zz}^{\text{PN}}(0, y \to 0^{\pm}) = \mp \frac{\mu b_x \nu}{\pi(1 - \nu)\zeta}. \tag{58}$$

For $d = a$ and $\zeta = a/[2(1 - \nu)]$, the jumps are

$$\sigma_{xx}^{\text{PN}}(0, y \to 0^{\pm}) = \mp \frac{2\mu b_x}{\pi a}, \quad \sigma_{zz}^{\text{PN}}(0, y \to 0^{\pm}) = \mp \frac{2\mu b_x \nu}{\pi a}. \tag{59}$$

The stresses $\sigma_{xy}^{\text{PN}}$ and $\sigma_{yy}^{\text{PN}}$ have the following extreme values: $|\sigma_{xy}^{\text{PN}}(x, 0)| = \frac{\mu b_x}{4\pi(1-\nu)\zeta}$ at $|x| = \zeta$ and $|\sigma_{yy}^{\text{PN}}(0, y)| = \frac{\mu b_x}{8\pi(1-\nu)\zeta}$ at $|y| = \zeta$. For W with $\nu = 0.28$ [63]: $\zeta \simeq 0.7a$ and $1/\zeta \simeq 1.44/a$, and $|\sigma_{xy}^{\text{PN}}(x, 0)| = \frac{\mu b_x}{2\pi a}$ at $|x| \simeq 0.7a$ and $|\sigma_{yy}^{\text{PN}}(0, y)| = \frac{\mu b_x}{8\pi a}$ at $|y| \simeq 0.7a$. Moreover, the stresses $\sigma_{xx}^{\text{PN}}$ and $\sigma_{zz}^{\text{PN}}$ become very large in the dislocation core region and at the dislocation line (see Fig. 9). The stresses are plotted in Fig. 10 for $d = a$.

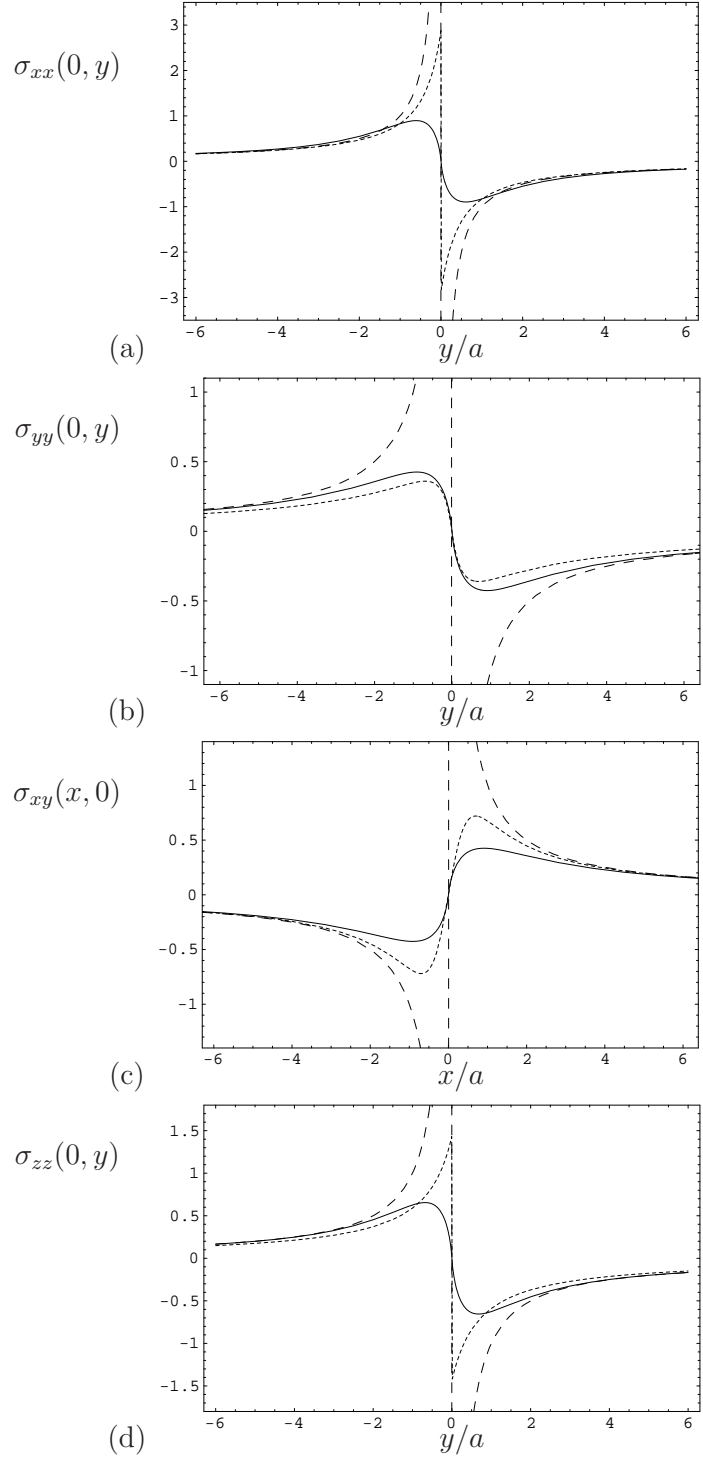

Figure 9: Stress components of an edge dislocation: (a) $\sigma_{xx}(0,y)$, (b) $\sigma_{yy}(0,y)$, (c) $\sigma_{xy}(x,0)$ are given in units of $\mu b_x/[2\pi(1-\nu)a]$ and (d) $\sigma_{zz}(0,y)$ is given in units of $\mu b_x \nu/[\pi(1-\nu)a]$ for strain gradient elasticity model for $\ell = 0.61a$ (solid line), Peierls-Nabarro model for $d = a$ (small dashed line), and Volterra model (dashed line) with $\nu = 0.28$.

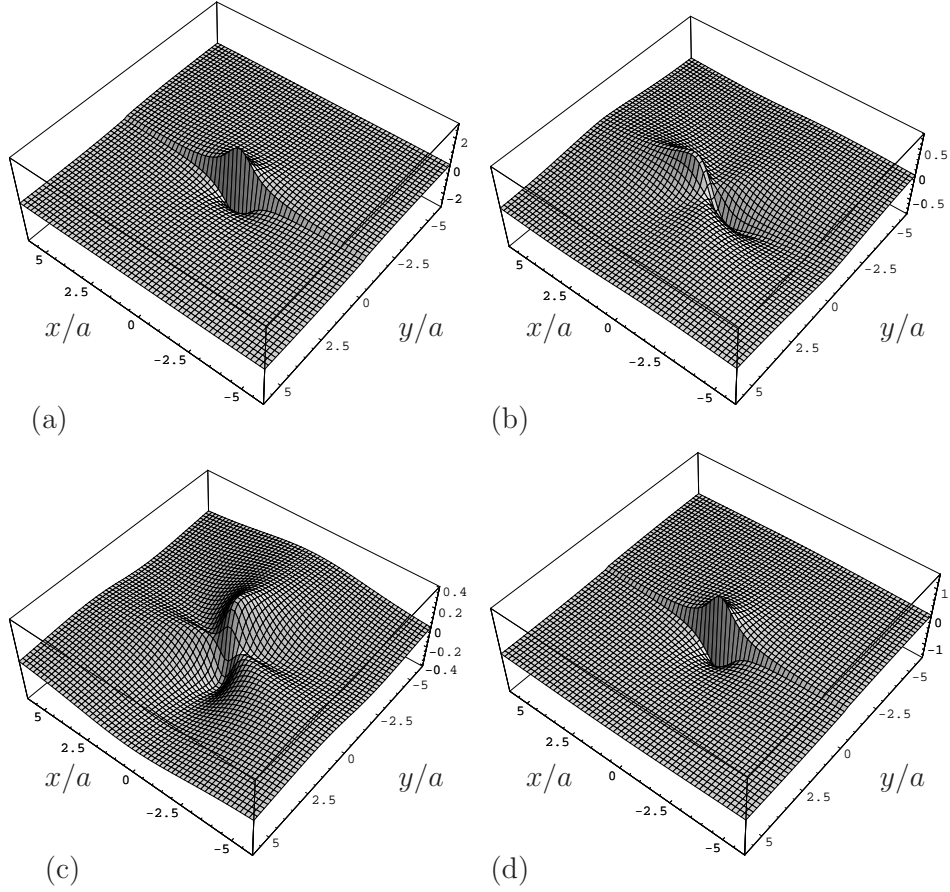

Figure 10: Stress fields of a Peierls-Nabarro edge dislocation for $d = a$: (a) $\sigma_{xx}$, (b) $\sigma_{xy}$, (c) $\sigma_{yy}$ are given in units of $\mu b_x/[2\pi(1-\nu)a]$ and (d) $\sigma_{zz}$ is given in units of $\mu b_x \nu/[\pi(1-\nu)a]$.

Using again the Eshelby smearing (ramp-core) technique [12], we replace the $\delta(x)$-function in the classical plastic distortion (47) by the $\delta$-sequence $f_\zeta(x)$:

$$f_\zeta(x) = \frac{1}{\pi} \frac{\zeta}{x^2 + \zeta^2} \tag{60}$$

and obtain for the plastic distortion of a Peierls-Nabarro edge dislocation

$$\beta_{xx}^{\mathrm{P,PN}} = \beta_{xx}^{\mathrm{P,0}} * f_\zeta = -b_x f_\zeta(x) H(-y) = -\frac{b_x}{\pi} \frac{\zeta}{x^2 + \zeta^2} H(-y). \tag{61}$$

The plastic distortion (61) possesses a minimum at the dislocation line

$$\beta_{xx}^{\mathrm{P,PN}}(0, y < 0) = -\frac{b_x}{\pi \zeta}. \tag{62}$$

The corresponding dislocation density of a Peierls-Nabarro edge dislocation reads

$$\alpha_{xz}^{\mathrm{PN}} = \partial_y \beta_{xx}^{\mathrm{P,PN}} = b_x f_\zeta(x) \delta(y) = \frac{b_x}{\pi} \frac{\zeta}{x^2 + \zeta^2} \delta(y). \tag{63}$$

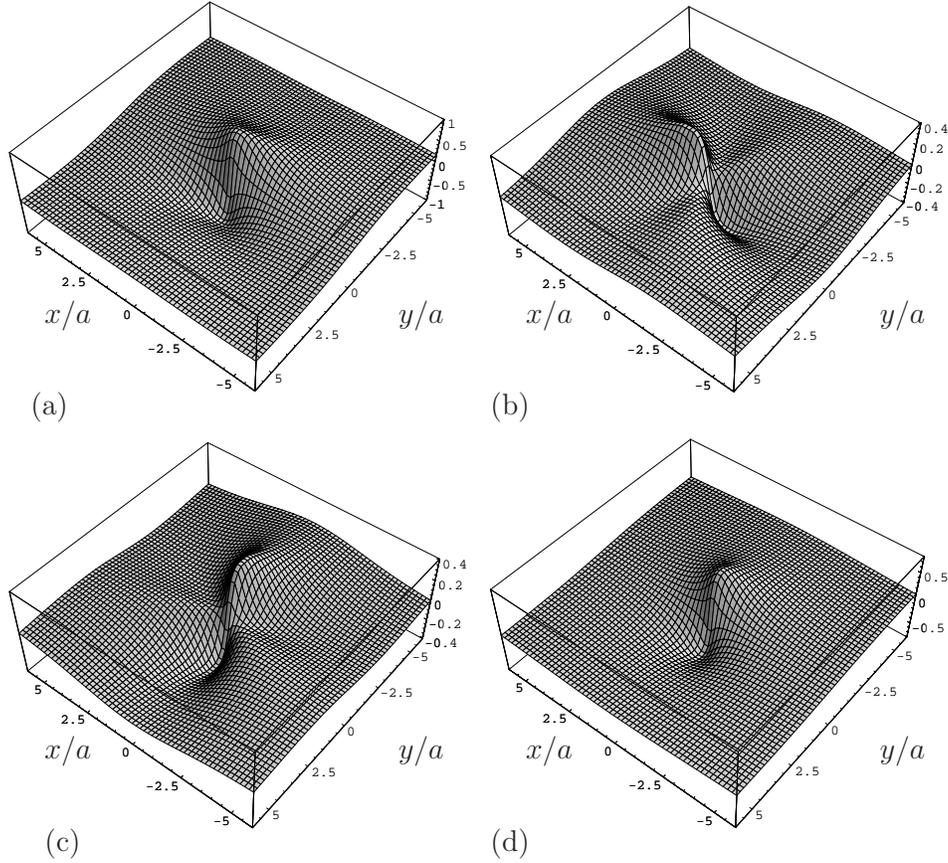

Figure 11: Stress fields of an edge dislocation in strain gradient elasticity for $\ell = 0.61a$: (a) $\sigma_{xx}$, (b) $\sigma_{xy}$, (c) $\sigma_{yy}$ are given in units of $\mu b_x/[2\pi(1-\nu)a]$ and (d) $\sigma_{zz}$ is given in units of $\mu b_x \nu/[\pi(1-\nu)a]$.

It can be seen that the relation $\alpha_{xz}^{\rm PN} = \alpha_{xz}^0 * f_\zeta$ is fulfilled. In Eq. (63), it can be seen that the dislocation density of a Peierls-Nabarro edge dislocation is still singular, due to the Delta-function $\delta(y)$.

### 3.3 Gradient solution

Let us now consider the corresponding fields of an edge dislocation within the theory of strain gradient elasticity of Helmholtz type given by Lazar and Maugin [30, 32, 33]. In strain gradient elasticity, the displacement fields of an edge dislocation are [32, 33]

$$u_x = \frac{b_x}{4\pi(1-\nu)}\left\{2(1-\nu)w(x,y) + \frac{xy}{r^2} - 4\ell^2\frac{xy}{r^4} + \frac{2xy}{r^2}K_2(r/\ell)\right\}, \tag{64}$$

$$u_y = -\frac{b_x}{4\pi(1-\nu)}\left\{(1-2\nu)\bigl(\ln r + K_0(r/\ell)\bigr) - \frac{y^2}{r^2} - 2\ell^2\frac{x^2-y^2}{r^4} + \frac{x^2-y^2}{r^2}K_2(r/\ell)\right\}, \tag{65}$$

where the displacement profile function $w(x, y)$ is given in Eq. (24). The displacement fields (64) and (65) are plotted in Fig. 8. Both displacement fields are non-singular and the field (64) possesses a discontinuity due to the dislocation profile function $w(x, y)$.

The incompatible elastic distortion of an edge dislocation is given by [32, 28]

$$\beta_{xx} = -\frac{b_x}{4\pi(1-\nu)} \frac{y}{r^2} \left\{ (1-2\nu) + \frac{2x^2}{r^2} + \frac{4\ell^2}{r^4}(y^2 - 3x^2) - \frac{2(y^2 - 3x^2)}{r^2} K_2(r/\ell) \right.$$
$$\left. - \frac{2(y^2 - \nu r^2)}{\ell r} K_1(r/\ell) \right\}, \qquad (66)$$

$$\beta_{xy} = \frac{b_x}{4\pi(1-\nu)} \frac{x}{r^2} \left\{ (3-2\nu) - \frac{2y^2}{r^2} - \frac{4\ell^2}{r^4}(x^2 - 3y^2) + \frac{2(x^2 - 3y^2)}{r^2} K_2(r/\ell) \right.$$
$$\left. - \frac{2(y^2 + (1-\nu)r^2)}{\ell r} K_1(r/\ell) \right\}, \qquad (67)$$

$$\beta_{yx} = -\frac{b_x}{4\pi(1-\nu)} \frac{x}{r^2} \left\{ (1-2\nu) + \frac{2y^2}{r^2} + \frac{4\ell^2}{r^4}(x^2 - 3y^2) - \frac{2(x^2 - 3y^2)}{r^2} K_2(r/\ell) \right.$$
$$\left. + \frac{2(y^2 - (1-\nu)r^2)}{\ell r} K_1(r/\ell) \right\}, \qquad (68)$$

$$\beta_{yy} = -\frac{b_x}{4\pi(1-\nu)} \frac{y}{r^2} \left\{ (1-2\nu) - \frac{2x^2}{r^2} - \frac{4\ell^2}{r^4}(y^2 - 3x^2) + \frac{2(y^2 - 3x^2)}{r^2} K_2(r/\ell) \right.$$
$$\left. - \frac{2(x^2 - \nu r^2)}{\ell r} K_1(r/\ell) \right\}. \qquad (69)$$

The stress components of an edge dislocation in strain gradient elasticity are [30, 15]

$$\sigma_{xx} = -\frac{\mu b_x}{2\pi(1-\nu)} \frac{y}{r^4} \left\{ (y^2 + 3x^2) + \frac{4\ell^2}{r^2}(y^2 - 3x^2) - 2y^2 \frac{r}{\ell} K_1(r/\ell) \right.$$
$$\left. - 2(y^2 - 3x^2) K_2(r/\ell) \right\}, \qquad (70)$$

$$\sigma_{yy} = -\frac{\mu b_x}{2\pi(1-\nu)} \frac{y}{r^4} \left\{ (y^2 - x^2) - \frac{4\ell^2}{r^2}(y^2 - 3x^2) - 2x^2 \frac{r}{\ell} K_1(r/\ell) \right.$$
$$\left. + 2(y^2 - 3x^2) K_2(r/\ell) \right\}, \qquad (71)$$

$$\sigma_{xy} = \frac{\mu b_x}{2\pi(1-\nu)} \frac{x}{r^4} \left\{ (x^2 - y^2) - \frac{4\ell^2}{r^2}(x^2 - 3y^2) - 2y^2 \frac{r}{\ell} K_1(r/\ell) \right.$$
$$\left. + 2(x^2 - 3y^2) K_2(r/\ell) \right\}, \qquad (72)$$

$$\sigma_{zz} = -\frac{\mu b_x \nu}{\pi(1-\nu)} \frac{y}{r^2} \left\{ 1 - \frac{r}{\ell} K_1(r/\ell) \right\}. \qquad (73)$$

The stresses (70)–(73) are non-singular and continuous. In addition, they are zero at the dislocation line. In fact, the "classical" singularities are eliminated. The stresses (70)–(73) have the following extreme values: $|\sigma_{xx}(0,y)| \simeq 0.546 \frac{\mu b_x}{2\pi(1-\nu)\ell}$ at $|y| \simeq 0.996\ell$, $|\sigma_{yy}(0,y)| \simeq 0.260 \frac{\mu b_x}{2\pi(1-\nu)\ell}$ at $|y| \simeq 1.494\ell$, $|\sigma_{xy}(x,0)| \simeq 0.260 \frac{\mu b_x}{2\pi(1-\nu)\ell}$ at $|x| \simeq 1.494\ell$, and $|\sigma_{zz}(0,y)| \simeq 0.399 \frac{\mu b_x \nu}{\pi(1-\nu)\ell}$ at $|y| \simeq 1.114\ell$. For W with $\ell = 0.61a$: $|\sigma_{xx}(0,y)| \simeq 0.895 \frac{\mu b_x}{2\pi(1-\nu)a}$ at $|y| \simeq 0.61a$, $|\sigma_{yy}(0,y)| \simeq 0.426 \frac{\mu b_x}{2\pi(1-\nu)a}$ at $|y| \simeq 0.91a$, $|\sigma_{xy}(x,0)| \simeq 0.426 \frac{\mu b_x}{2\pi(1-\nu)\ell}$ at $|x| \simeq 0.91a$, and $|\sigma_{zz}(0,y)| \simeq 0.654 \frac{\mu b_x \nu}{\pi(1-\nu)a}$ at $|y| \simeq 0.68a$. The non-singular stresses are plotted in Figs. 9 and 10.

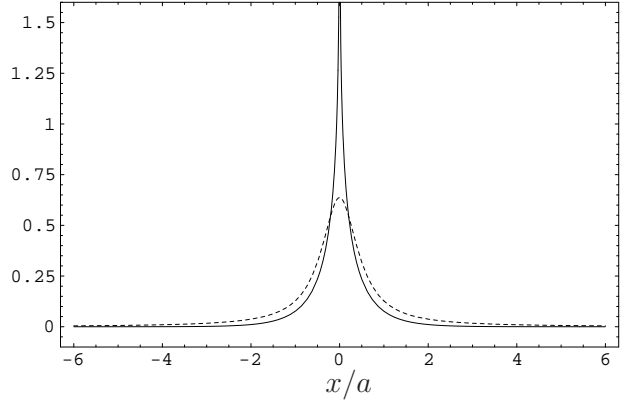

Figure 12: Core spreading function (dislocation density) in strain gradient elasticity for $\ell = 0.61a$ (solid line) and Peierls-Nabarro model for $\eta = a/2$ (small dashed line).

The plastic distortion and the dislocation density of an edge dislocation in strain gradient elasticity are [32]

$$\beta_{xx}^{\mathrm{P}} = -\frac{b_x}{2\pi} \int_0^\infty \frac{\cos(sx)}{1+\ell^2 s^2} \left[\operatorname{sgn}(y)\, \mathrm{e}^{-|y|\sqrt{s^2+\frac{1}{\ell^2}}} + 2H(-y)\right] \mathrm{d}s, \tag{74}$$

$$\alpha_{xz} = \frac{b_x}{2\pi\ell^2}\, K_0(r/\ell). \tag{75}$$

Such a two-dimensional dislocation core shape is in agreement with the measurement of the dislocation core distribution by digital processing of high transmission electron microscopy micrographs [21]. Moreover, the dislocation density tensor (75) obtained in strain gradient elasticity describes a dislocation core spreading in agreement with atomistic modeling [17].

## 4 Conclusions

In this paper, we have compared the Peierls-Nabarro dislocation continuum model with the strain gradient elasticity dislocation model. We have shown that the strain gradient elasticity dislocation model represents a more realistic dislocation continuum model than the Peierls-Nabarro model of dislocations.

First of all, the non-singular stresses of dislocations in strain gradient elasticity are not so large in the dislocation core and at the dislocation line as the non-singular stresses of the Peierls-Nabarro dislocations. Second, the non-singular stresses of dislocations in strain gradient elasticity are everywhere continuous and they are zero at the dislocation line, whereas some components of the non-singular stresses of the Peierls-Nabarro dislocations possess discontinuities (jumps) across the $y = 0$ plane. Nonetheless, for the Peierls-Nabarro model the stresses as well as the elastic strains are very large in order to be realistic in the core region. The profile (in $x$-direction) of the displacement jump of a Peierls-Nabarro dislocation is wider than that of a dislocation in the strain gradient elasticity dislocation model. Using the displacement profile of dislocations, strain gradient elasticity leads to a narrower dislocation core than the core of a Peierls-Nabarro dislocation. In addition, strain gradient elasticity is a non-singular dislocation continuum theory including weak nonlocality in the dislocation core region. In the Peierls-Nabarro dislocation model, nonlinearity is present across the $y = 0$ plane, whereas in the strain gradient elasticity dislocation model, nonlocality is present in the dislocation core region. In the Peierls-Nabarro dislocation model, the dislocation core has the shape of a line in $x$-direction (flat core or ramp core) in the $y = 0$ plane, whereas the strain gradient elasticity dislocation model, the dislocation core has the shape of a disk (isotropic core) in the $xy$-plane. From that point of view, strain gradient elasticity is able to model a more realistic two-dimensional dislocation core than the one-dimensional ramp-core or flat core of the Peierls-Nabarro dislocations. The size of the dislocation core is only a few atoms wide (see Fig. 12).

In the Peierls-Nabarro model of straight dislocations, the regularization function, which has the physical meaning of the dislocation core spreading function is a one-dimensional delta sequence (Lorentzian-type) $f_\varepsilon(x)$, which is a continuous distribution on a line known from the theory of generalized functions [14, 19]. The regularization of the dislocation fields in the Peierls-Nabarro model of straight dislocations is given by (see also [23, 67, 18, 42, 52])

$$u_i^{\text{PN}}(x,y) = u_i^0(x,y) * f_\varepsilon(x), \tag{76}$$

$$\beta_{ij}^{\text{PN}}(x,y) = \beta_{ij}^0(x,y) * f_\varepsilon(x), \tag{77}$$

$$\sigma_{ij}^{\text{PN}}(x,y) = \sigma_{ij}^0(x,y) * f_\varepsilon(x), \tag{78}$$

$$\beta_{ij}^{\text{P,PN}}(x,y) = \beta_{ij}^{\text{P,0}}(x,y) * f_\varepsilon(x), \tag{79}$$

$$\alpha_{ij}^{\text{PN}}(x,y) = \alpha_{ij}^0(x,y) * f_\varepsilon(x), \tag{80}$$

where $*$ denotes the one-dimensional convolution in the variable $x$. For a screw dislocation $\varepsilon = \eta$ and for an edge dislocation $\varepsilon = \zeta$. However, in the variable core model [42, 43] the half-size of the dislocation core should be considered as independent parameter unlike the original Peierls-Nabarro model where it was based on the assumption of the sinusoidal relationship between the shear stress and the slip discontinuity along the slip plane.

In the strain gradient elasticity model of straight dislocations, the regularization function which has the physical meaning of the dislocation core spreading function is the two-dimensional Green function of the (modified) Helmholtz operator $L$ [68, 65]

$$L\, G^L(x,y) = \delta(x)\delta(y), \quad \text{with} \quad L = 1 - \ell^2 \Delta, \tag{81}$$

where $\Delta$ is the Laplace operator. $G^L(x,y)$ is also a delta sequence. The regularization of the dislocation fields in the strain gradient elasticity theory of straight dislocations

Table 1: Comparison between the Peierls-Nabarro dislocation model and the strain gradient elasticity dislocation model for straight dislocations.

|  | Peierls-Nabarro model | Gradient elasticity model |
|---|---|---|
| physical motivation | nonlinearity | nonlocality |
| regularization | in $x$-direction | isotropic in the $xy$-plane |
| core spreading | $f_\varepsilon(x) = \dfrac{1}{\pi}\dfrac{\varepsilon}{x^2+\varepsilon^2}$ | $G^L(x,y) = \dfrac{1}{2\pi\ell^2}\,K_0(r/\ell)$ |
| function | Lorentzian | modified Bessel function $K_0$ |
| core parameter | $\varepsilon > 0$ | $\ell > 0$ |
| core shape | flat core (1D) | isotropic (2D) |
| classical limit | $\lim\limits_{\varepsilon\to 0} f_\varepsilon(x) = \delta(x)$ | $\lim\limits_{\ell\to 0} G^L(x,y) = \delta(x)\delta(y)$ |
| normalization | $\int_{\mathbb{R}} f_\varepsilon(x)\,\mathrm{d}x = 1$ | $\int_{\mathbb{R}^2} G^L(x,y)\,\mathrm{d}x\,\mathrm{d}y = 1$ |
| dislocation density | $\alpha_{iz}^{\mathrm{PN}} = b_i\, f_\varepsilon(x)\delta(y)$ | $\alpha_{iz} = b_i\, G^L(x,y)$ |
| stress | non-singular, discontinuous | non-singular, continuous |
| displacement | non-singular, discontinuous | non-singular, discontinuous |

is given by (see [32, 28, 29])

$$u_i(x,y) = u_i^0(x,y) * G^L(x,y)\,, \qquad (82)$$

$$\beta_{ij}(x,y) = \beta_{ij}^0(x,y) * G^L(x,y)\,, \qquad (83)$$

$$\sigma_{ij}(x,y) = \sigma_{ij}^0(x,y) * G^L(x,y)\,, \qquad (84)$$

$$\beta_{ij}^{\mathrm{P}}(x,y) = \beta_{ij}^{\mathrm{P},0}(x,y) * G^L(x,y)\,, \qquad (85)$$

$$\alpha_{ij}(x,y) = \alpha_{ij}^0(x,y) * G^L(x,y)\,, \qquad (86)$$

where $*$ denotes the two-dimensional convolution in $x$ and $y$.

A comparison between the Peierls-Nabarro dislocation model and the strain gradient elasticity dislocation model is given in Table 1. It becomes evident that the strain gradient elasticity dislocation model represents a better and more physical dislocation continuum model than the Peierls-Nabarro dislocation model.

## Acknowledgement

The author gratefully acknowledges Dr. Eleni Agiasofitou for constructive remarks, which significantly influenced this work. Remarks from Prof. Xanthippi Markenscoff concerning the variable core model are gratefully acknowledged. The author also wishes to thank two anonymous reviewers for their comments. The author acknowledges a grant from the Deutsche Forschungsgemeinschaft (Grant No. La1974/4-1).

# A  Strain gradient elasticity theory of Helmholtz type

In this Appendix, we review the basics of the theory of strain gradient elasticity of Helmholtz type given by Lazar and Maugin [30], Lazar *et al.* [31], Lazar [28, 29]. In particular, we give the fundamentals of strain gradient elasticity of Helmholtz type including the equilibrium condition in terms of the Cauchy stress tensor and double stress tensor.

In isotropic gradient elasticity of Helmholtz type, the strain energy density is of the form

$$W = \frac{1}{2} C_{ijkl}\beta_{ij}\beta_{kl} + \frac{1}{2}\ell^2 C_{ijkl}\partial_m\beta_{ij}\partial_m\beta_{kl}\,, \tag{A.1}$$

where $C_{ijkl}$ is the isotropic tensor of the elastic moduli given by

$$C_{ijkl} = \mu\bigl(\delta_{ik}\delta_{jl} + \delta_{il}\delta_{jk}\bigr) + \lambda\,\delta_{ij}\delta_{kl}\,. \tag{A.2}$$

Here $\mu$ and $\lambda$ are the Lamé moduli, and $\ell$ is the characteristic length of gradient elasticity of Helmholtz type.

The Cauchy stress tensor reads

$$\sigma_{ij} = \frac{\partial W}{\partial e_{ij}} = C_{ijkl}\beta_{kl}\,, \tag{A.3}$$

where $e_{ij} = 1/2(\beta_{ij} + \beta_{ji})$ is the elastic strain tensor and $\beta_{ij}$ is the elastic distortion tensor. The double stress tensor reads

$$\tau_{ijk} = \frac{\partial W}{\partial(\partial_k e_{ij})} = \ell^2\, C_{ijmn}\partial_k\beta_{mn} = \ell^2 \partial_k\sigma_{ij}\,. \tag{A.4}$$

The equilibrium condition (for vanishing body forces) in terms of the Cauchy stress tensor and double stress tensor is given by

$$\partial_j(\sigma_{ij} - \partial_k\tau_{ijk}) = 0\,. \tag{A.5}$$

In presence of defects such as dislocations, the total distortion tensor $\beta_{ij}^{\mathrm{T}}$ can be decomposed into an elastic distortion part $\beta_{ij}$ and a plastic distortion part $\beta_{ij}^{\mathrm{P}}$ as follows

$$\beta_{ij}^{\mathrm{T}} := \partial_j u_i = \beta_{ij} + \beta_{ij}^{\mathrm{P}}\,, \tag{A.6}$$

where $u_i$ denotes the displacement vector. Since dislocations cause self-stresses, body forces are zero. The dislocation density tensor can be defined in terms of the plastic and elastic distortion tensors as follows (e.g., [22])

$$\alpha_{ij} = -\epsilon_{jkl}\partial_k\beta_{il}^{\mathrm{P}}\,, \quad \text{or} \quad \alpha_{ij} = \epsilon_{jkl}\partial_k\beta_{il}\,. \tag{A.7}$$

The dislocation density tensor satisfies the dislocation Bianchi identity

$$\partial_j\alpha_{ij} = 0\,, \tag{A.8}$$

which means that a dislocation cannot end inside the medium. Then the equilibrium condition reads in terms of the displacement vector and the plastic distortion tensor

$$LL_{ik}u_k = C_{ijkl}\partial_j L\beta^{\text{P}}_{kl}\,, \tag{A.9}$$

where $L = 1 - \ell^2\Delta$ is the (modified) Helmholtz operator and $L_{ik} = C_{ijkl}\partial_j\partial_l$ is the Navier operator. Using the two inhomogeneous Helmholtz equations

$$Lu_k = u_k^0\,, \tag{A.10}$$

$$L\beta^{\text{P}}_{kl} = \beta^{\text{P},0}_{kl}\,, \tag{A.11}$$

we obtain the Navier equation known from classical eigenstrain theory [48]

$$L_{ik}u_k^0 = C_{ijkl}\partial_j\beta^{\text{P},0}_{kl}\,. \tag{A.12}$$

Therefore, the fields $u_k^0$ and $\beta_{kl}^0$ may be identified with the classical displacement vector and classical plastic distortion tensor of classical incompatible elasticity theory of dislocations. Finally, using Eqs. (A.9) and (A.11), the displacement vector $u_k$ is determined by the inhomogeneous Helmholtz-Navier equation [28, 29]

$$LL_{ik}u_k = C_{ijkl}\partial_j\beta^{\text{P},0}_{kl}\,, \tag{A.13}$$

where the right hand side is given by the gradient of the classical plastic distortion tensor (classical eigendistortion). In addition, the elastic distortion tensor $\beta_{km}$ satisfies the inhomogeneous Helmholtz-Navier equation [28, 29]

$$LL_{ik}\beta_{km} = -C_{ijkl}\epsilon_{mlr}\partial_j\alpha^0_{kr}\,, \tag{A.14}$$

where the right hand side is given by the gradient of the classical dislocation density tensor $\alpha^0_{kr}$.

# B  Characteristic lengths of strain gradient elasticity for some cubic materials

An important issue in strain gradient elasticity theory is the determination of the characteristic lengths in addition to the elastic constants. In Table 2, the characteristic lengths $\ell_1$ and $\ell_2$ of Mindlin's isotropic strain gradient elasticity theory [45] are reported for several fcc and bcc crystals determined by ab initio density functional theory (DFT) method [61]. From Mindlin's characteristic lengths $\ell_1$ and $\ell_2$, we can determine the corresponding characteristic length of gradient elasticity of Helmholtz type via (see Table 2)

$$\ell = \frac{\ell_1 + \ell_2}{2}\,. \tag{B.1}$$

Thus, the characteristic length $\ell$ is the average of the two characteristic lengths $\ell_1$ and $\ell_2$ of Mindlin's isotropic gradient elasticity theory [60]. It is remarkable that for the three fcc-crystals, it holds $\ell/a \approx 0.5$, and for the three bcc-crystals, it holds $\ell/a \approx 0.6$ (see Table 2).

Table 2: Calculated characteristic lengths and equilibrium lattice parameter for several fcc and bcc crystals via ab initio [61].

| Material | Crystal | $\ell_1$(Å) | $\ell_2$(Å) | $\ell = \dfrac{\ell_1 + \ell_2}{2}$(Å) | $a$(Å) | $\ell/a$ |
|---|---|---|---|---|---|---|
| Ir | fcc | 2.1523 | 1.8217 | 1.9870 | 3.87 | 0.51 |
| Pt | fcc | 2.4480 | 1.6353 | 2.0416 | 3.92 | 0.52 |
| Al | fcc | 2.3415 | 1.6582 | 1.9998 | 4.05 | 0.49 |
| W  | bcc | 1.6460 | 2.2026 | 1.9243 | 3.15 | 0.61 |
| V  | bcc | 1.5519 | 2.1710 | 1.8614 | 3.02 | 0.62 |
| Mo | bcc | 1.6380 | 2.2438 | 1.9409 | 3.16 | 0.61 |

Only a few cubic crystals such as tungsten (W) and aluminum (Al) are elastically isotropic or nearly isotropic materials [6]. Thus, tungsten (W) is a proper material to test the theory of isotropic strain gradient elasticity of Helmholtz type and is used for the numerical analysis in this work.

# References

[1] N.C. Admal, J. Marian, and G. Po, *The atomistic representation of first strain-gradient elastic tensors*, J. Mech. Phys. Solids 99 (2017), 93–115.

[2] M. Born and K. Huang, *Dynamical Theory of Crystal Lattices*, Clarendon Press, Oxford, 1954.

[3] V.V. Bulatov and W. Cai, *Computer Simulations of Dislocations*, Oxford University Press, Oxford, 2006.

[4] W. Cai, A. Arsenlis, C.R. Weinberger, and V.V. Bulatov, *A non-singular continuum theory of dislocations*, J. Mech. Phys. Solids 54 (2006), 561–587.

[5] K.M. Davoudi, H.M. Davoudi, and E.C. Aifantis, *Nanomechanics of a screw dislocation in a functionally graded material using the theory of gradient elasticity*, J. Mech. Behav. Mater. 21 (2013), 187–194.

[6] P.H. Dederichs and G. Leibfried, *Elastic Green's function for anisotropic cubic crystals*, Phys. Rev. 188 (1969), 1175–1183.

[7] R. deWit, *Theory of disclinations IV*, J. Res. Nat. Bur. Stand. (U.S.) 77A (1973), 607–658.

[8] H.-D. Dietze, *Versetzungsstrukturen in kubisch-flächenzentrierten Kristallen. II*, Z. Phys. 131 (1952), 156–169.

[9] D.G.B. Edelen, *A correct, globally defined solution of the screw dislocation problem in the gauge theory of defects*, Int. J. Engng. Sci. 34 (1996), 81–86.

[10] T. Egami, K. Maeda, and V. Vitek, *Structural defects in amorphous solids: A computer simulation study*, Phil. Mag. A 41 (1980), 883–901.

[11] A.C. Eringen, *Nonlocal Continuum Field Theories*, Springer, New York, 2002.

[12] J.D. Eshelby, *Lecture on the elastic energy-momentum tensor*, 1977, in: Lecture Notes (handwritten), *Collected Works of J.D. Eshelby*, Markenscoff, X., Gupta, A. (Eds.), Springer, 2006.

[13] D. Ferré, P. Carrez, and P. Cordier, *Modeling dislocation cores in $SrTiO_3$ using the Peierls-Nabarro model*, Phys. Rev. B 77 (2008), 014106.

[14] I.M. Gel'fand and G.E. Shilov, *Generalized Functions, Vol. I*, Academic, New York 1964.

[15] M.Yu. Gutkin and E.C. Aifantis, *Dislocations in gradient elasticity*, Scripta Mater. 40 (1999), 559–566.

[16] M.Yu. Gutkin, *Elastic behaviour of defects in nanomaterials I*, Rev. Adv. Mater. Sci. 13 (2006), 125–161.

[17] C.S. Hartley and Y. Mishin, *Characterization and visualization of the lattice misfit associated with dislocation cores*, Acta Mater. 53 (2005), 1313–1321.

[18] J.P. Hirth and J. Lothe, *Theory of Dislocations*, Wiley, New York, 1968.

[19] R.P. Kanwal, *Generalized Functions: Theory and Applications*, 3rd ed., Birkhäuser, Boston, 2004.

[20] H. Koizumi, H.O.K. Kirchner, and T. Suzuki, *Construction of the Peierls-Nabarro potential of a dislocation from interatomic potentials*, Phil. Mag. 86 (2006), 3835–3846.

[21] S. Kret, P. Dłużewski, P. Dłużewski, and W. Sobczak, *Measurement of the dislocation core distribution by digital processing of high transmission electron microscopy micrographs: a new technique for studying defects*, J. Phys.: Condens. Matter 12 (2000), 10313–10318.

[22] E. Kröner, *Kontinuumstheorie der Versetzungen und Eigenspannungen*, Springer, Berlin, 1958.

[23] R.W. Lardner, *Mathematical Theory of Dislocations and Fracture*, University of Toronto Press, Toronto, 1974.

[24] M. Lazar, *An elastoplastic theory of dislocations as a physical field theory with torsion*, J. Phys. A: Math. Gen. 35 (2002), 1983–2004.

[25] M. Lazar, *A nonsingular solution of the edge dislocation in the gauge theory of dislocations*, J. Phys. A: Math. Gen. 36 (2003), 1415–1437.

[26] M. Lazar, *Dislocations in the field theory of elastoplasticity*, Comput. Mater. Sci. 28 (2003), 419–428.

[27] M. Lazar, *Non-singular dislocation loops in gradient elasticity*, Phys. Lett. A 376 (2012), 1757–1758.

[28] M. Lazar, *The fundamentals of non-singular dislocations in the theory of gradient elasticity: Dislocation loops and straight dislocations*, Int. J. Solids Struct. 50 (2013), 352–362.

[29] M. Lazar, *On gradient field theories: gradient magnetostatics and gradient elasticity*, Phil. Mag. 94 (2014), 2840–2874.

[30] M. Lazar and G.A. Maugin, *Nonsingular stress and strain fields of dislocations and disclinations in first strain gradient elasticity*, Int. J. Engng. Sci. 43 (2005), 1157–1184.

[31] M. Lazar, G.A. Maugin, and E.C. Aifantis, *On dislocations in a special class of generalized elasticity*, Phys. Status Solidi (b) 242 (2005), 2365–2390.

[32] M. Lazar and G.A. Maugin *Dislocations in gradient elasticity revisited*, Proc. R. Soc. Lond. A 462 (2006), 3465–3480.

[33] M. Lazar and G.A. Maugin, *A note on line forces in gradient elasticity*, Mech. Res. Commun. 33 (2006), 674–680.

[34] M. Lazar, G.A. Maugin, and E.C. Aifantis, *Dislocations in second strain gradient elasticity*, Int. J. Solids Struct. 43 (2006), 1787–1817.

[35] M. Lazar, E. Agiasofitou, and D. Polyzos, *On gradient enriched elasticity theories: A reply to "Comment on 'On non-singular crack fields in Helmholtz type enriched elasticity theories' " and important theoretical aspects*, Eprint Archive: https://arxiv.org/abs/1504.00869, (2015).

[36] G. Leibfried and K. Lücke, *Über das Spannungsfeld einer Versetzung*, Z. Phys. 126 (1949), 450–464.

[37] G. Leibfried and H.-D. Dietze, *Zur Theorie der Schraubenversetzung*, Z. Phys. 126 (1949), 790–808.

[38] G. Leibfried and H.-D. Dietze, *Versetzungsstrukturen in kubisch-flächenzentrierten Kristallen. I*, Z. Phys. 131 (1951), 113–129.

[39] C.L. Lee and S. Li, *A half-space Peierls-Nabarro model and the mobility of screw dislocations in a thin film*, Acta Mater. 55 (2007), 2149–2157.

[40] C.L. Lee and S. Li, *The size effect of thin films on the Peierls stress of edge dislocations*, Math. Mech. Solids 13 (2008), 316–335.

[41] S. Li and G. Wang, *Introduction to Micromechanics and Nanomechanics*, World Scientific, Singapore, 2008.

[42] V.A. Lubarda, X. Markenscoff, *Variable core model and the Peierls stress for the mixed (screw-edge) dislocation*, Appl. Phys. Lett. 89 (2006), 151923.

[43] V.A. Lubarda, X. Markenscoff, *Configurational force on a lattice dislocation and the Peierls stress*, Arch. Appl. Mech. 77 (2007), 147–154.

[44] R. Miller, R. Phillips, G. Beltz, and M. Ortiz, *A non-local formulation of the Peierls dislocation model*, J. Mech. Phys. Solids 46 (1998), 1845–1867.

[45] R.D. Mindlin, *Micro-structure in linear elasticity*, Arch. Rat. Mech. Anal. 16 (1964), 51–78.

[46] R.D. Mindlin, *Second gradient of strain and surface-tension in linear elasticity*, Int. J. Solids Struct. 1 (1965), 417–438.

[47] R.D. Mindlin and N.N. Eshel, *On first strain gradient theory in linear elasticity*, Int. J. Solids Struct. 4 (1968), 109–124.

[48] T. Mura, *Micromechanics of Defects in Solids*, 2nd ed., Martinus Nijhoff, Dordrecht, 1987.

[49] F.R.N. Nabarro, *Dislocations in a simple cubic lattice*, Proc. Phys. Soc. (London) 59 (1947), 256–272.

[50] F.R.N. Nabarro, *Mathematical theory of stationary dislocations*, Adv. Phys. 1 (1952), 269–394.

[51] F.R.N. Nabarro, *Theory of Crystal Dislocations*, Oxford University Press, Oxford, 1967.

[52] L. Ni and X. Markenscoff, *The self-force and effective mass of a generally accelerating dislocation I: Screw dislocation*, J. Mech. Phys. Solids 56 (2008), 1348-1379.

[53] R.E. Peierls, *The size of a dislocation*, Proc. Phys. Soc. (London) 54 (1940), 34-37.

[54] R.C. Picu, *The Peierls stress in non-local elasticity*, J. Mech. Phys. Solids 50 (2002), 717–735.

[55] G. Po, M. Lazar, D. Seif, and N. Ghoniem, *Singularity-free dislocation dynamics with strain gradient elasticity*, J. Mech. Phys. Solids 68 (2014), 161–178.

[56] G. Po, M. Lazar, N.C. Admal, and N. Ghoniem, *A non-singular theory of dislocations in anisotropic crystals*, Eprint Archive: https://arxiv.org/abs/1706.00828, (2017).

[57] G. Schoeck, *The core structure of dislocations in Al: a critical assessment*, Mater. Sci. Eng. A. 333 (2002), 390–396.

[58] G. Schoeck, *Atomic dislocation core parameters*, Phys. Status Solidi (b) 247 (2010), 265–268.

[59] A. Seeger, *Theorie der Gitterfehlstellen*, in: Handbuch der Physik VII/1 (ed. S. Flügge), Springer, Berlin (1955), pp. 383–665.

[60] H.M. Shodja and A. Tehranchi, *A formulation for the characteristic lengths of fcc materials in first strain gradient elasticity via Sutton-Chen potential*, Phil. Mag. 90 (2010), 1893–1913; *Corrigendum*, Phil. Mag. 92 (2012), 1170–1171.

[61] H.M. Shodja, A. Zaheri, and A. Tehranchi, *Ab initio calculations of characteristic lengths of crystalline materials in first strain gradient elasticity*, Mech. Mater. 61 (2013), 73–78.

[62] E.B. Tadmor and R.E. Miller, *Modeling Materials: Continuum, Atomistic and Multiscale Techniques*, Cambridge University Press, Cambridge, 2011.

[63] C. Teodosiu, *Elastic Models of Crystal Defects*, Springer, Berlin, 1982.

[64] R.A. Toupin and D.C. Grazis, *Surface effects and initial stress in continuum and lattice models of elastic crystals*, in: Proceedings of the International Conference on Lattice Dynamics, Copenhagen. Edited by R.F. Wallis, Pergamon Press (1964), pp. 597–602.

[65] V.S. Vladimirow, *Equations of Mathematical Physics*, Dekker, New York, 1971.

[66] A.M. Walker, P. Carrez, and P. Cordier, *Atomic-scale models of dislocation cores in minerals: progress and prospects*, Mineral. Mag. 74 (2010), 381–413.

[67] J. Weertman and J. R. Weertman, *Moving dislocations*, in: Nabarro, F.R.N (Ed.), Dislocations in Solids, Vol. 3. North-Holland, Amsterdam (1980), pp. 1–59.

[68] E. Zauderer, *Partial Differential Equations of Applied Mathematics*, John Wiley & Sons Inc, New York, 1983.